\documentclass[aps,physrev,reprint]{revtex4-2}

\usepackage{graphicx}%
\usepackage{dcolumn}
\usepackage{bm}

\usepackage{url}
\usepackage[english]{babel}

\usepackage{multirow}

\usepackage{subfigure}

\usepackage{amsmath}
\usepackage{mathtools}
\usepackage{gensymb}
\usepackage{geometry}
\usepackage{booktabs}
\usepackage[utf8]{inputenc}
\usepackage{nicefrac}
\usepackage{array}

\usepackage{siunitx}
\usepackage[T1]{fontenc}
\usepackage{hyperref}
\usepackage[mathlines]{lineno}

\newcommand{\p}{$\%$}
\newcommand{\fn}{Fe$_4$N}

\begin{document}

\title{\textbf{Extending the Growth Temperature-N Concentration Regime Through Pd Doping in Fe$_4$N Thin Films}}

\author{Rohit Kumar Meena}
\affiliation{UGC-DAE Consortium for Scientific Research, University Campus, Khandwa Road, Indore, 452001, India}
 

\author{Akhil Tayal}
\affiliation{Brookhaven National Laboratory (BNL) National Synchrotron Light Source II (NSLS-II) 8-ID ISS beam line , New York, United States}

\author{Andrei Gloskovskii}
\affiliation{Deutsches Elektronen-Synchrotron, Notkestrasse 85, 22607 Hamburg, Germany}

\author{Mukul Gupta$^*$}
\affiliation{UGC-DAE Consortium for Scientific Research, University Campus, Khandwa Road, Indore, 452001, India}
\author {}
\address{$^*$ Corresponding author email: mgupta@csr.res.in}

\date{\today}

\begin{abstract} Fe$_4$N is a well-known anti-perovskite compound exhibiting high magnetization, high chemical stability, low coercivity, high Curie temperature, and high spin-polarization ratio.~Therefore, it is a viable candidate for applications in spintronic and magnetic storage devices.~However, the Fe$_4$N phase is formed in a narrow substrate temperature (Ts)-N concentration (Nc) regime in the phase diagram of Fe-N. It has been observed that a slight N deficiency will lead to impurity of $\alpha$-Fe, and some N efficiency would result in $\epsilon$-Fe$_3$N phase. Through this work, it has been demonstrated that the doping of Pd can be suitably utilized to extend the Ts-Nc regime for the growth of \fn~thin films. EXAFS analysis indicate that Pd atoms are subsituting corener Fe atoms. Magnetization measurements reveal that the saturation magnetization reduces nominally with Pd doping up to 13 at.\p. Therefore, it is foreseen that Pd doping is effective in extending the Fe$_4$N phase formation regime without a significant impact on its structural, electronic,  and magnetic properties.
\end{abstract}
\maketitle

\section{Introduction}
\begin{figure}
        \centering
        \includegraphics[width= 0.8 \linewidth ]{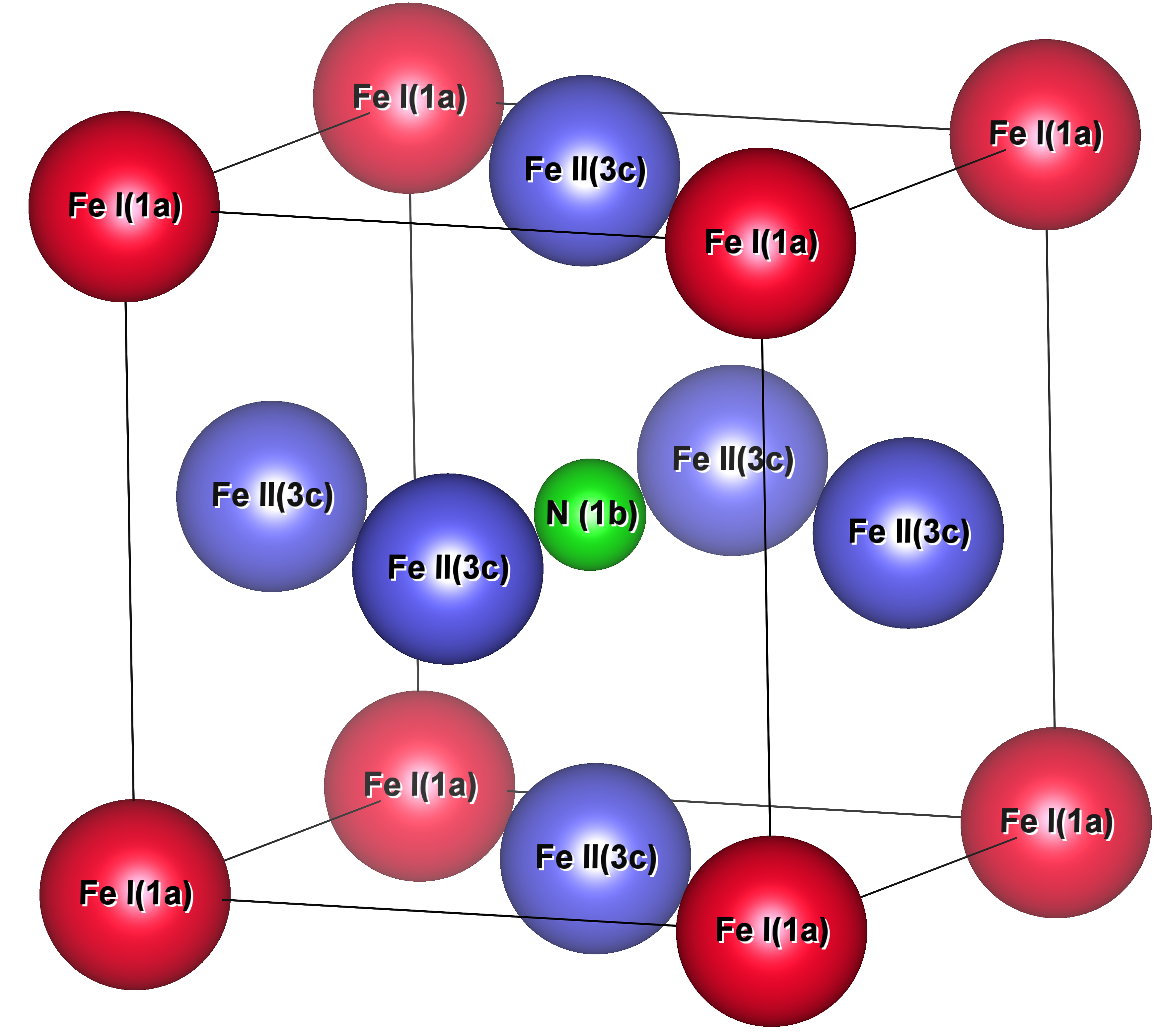}
    \caption{Crystal structure of Fe$_4$N. Fe atoms are shown in red, blue and N atom in green.}
    \label{crystal structure}
\end{figure}

Fe$_4$N have been studied extensively due to its high saturation magnetization (Ms) ($\approx$~2.45 $\mathrm{\mu_B}$/Fe atom) \cite{moment}, high chemical stability, low coercivity and high Curie temperature (Tc $ \approx$ 760\,K) \cite{Curietemp}. Additionally, its spin-polarization ratio (SPR) is predicted to reach unity~\cite{SPR_thoery}, while the experimental value obtained so far is 81.3\,\p~\cite{SPR}. Therefore, Fe$_4$N has been explored extensively for various applications such as spintronic and high density magnetic storage devices \cite{applications}.~Fe$_4$N crystallizes in a simple cubic anti-perovskite structure (space group: Pm$\Bar{3}$m) in a way that the N atom occupies the body center position in the fcc-Fe metal lattice as shown in figure~\ref{crystal structure}. The N insertion in the fcc-Fe metal lattice leads to the formation of two in-equivalent metal sites, one in the corner position (Fe I) and another in the face center position (Fe II), which results in an increased value of the Ms due to magneto-volume effect \cite{Curietemp,Akshaya}. Also, N insertion results in high SPR due to strong hybridization between 3$d$-2$p$ orbitals of Fe and N atoms \cite{SPR_thoery}.~The predicted and experimentally observed value of the lattice parameter (LP) for Fe$_4$N is 3.795\,\AA~\cite{LatticeP}, which is higher than that of hypothetical fcc-Fe (3.571\,\AA).

Thin films of Fe$_4$N are generally formed within a sharply defined temperature–nitrogen concentration (Ts- Nc) regime, centered at 20 at.\,\p\,and at a substrate temperature (Ts) $\approx$ 673\,K. A slight nitrogen deficiency introduces $\alpha$-Fe impurities, whereas a slight excess promotes the formation of the $\epsilon$-Fe$_3$N phase. Such deviations from the ideal \fn~composition often results in a discrepancy of the measured Ms values, typically ranging between 1.25 - 2.9\,$\mathrm{\mu_B}$/Fe atom \cite{High_Ms,Low_Ms}. In order to resolve this problem, few approaches have been adopted like single-phase \fn~ has been achieved by controlling film-substrate interdiffusion using Ag, Cu, and CrN under layers~\cite{buffer_layer}, employing epitaxial growth using different techniques~\cite{Interface}, and tuning growth parameters like deposition rates etc.~\cite{growthparameters}.

Through this work, we are proposing a new approach in which Pd doping can be utilized to stabilize the single phase \fn. We have varied the Pd doping concentration at 5, 13 and 24 at.\,\p~in \fn~which resulted into an extended (Ts-Nc) regime.~This substituted iron nitride is iso-structural to \fn~and prefer the replacement at the 1a wyckoff site~\cite{substitute1}. Although, Pd-doped Fe$_4$N has been investigated predominantly through theoretical studies focusing on its electronic, magnetic, and elastic properties, experimental works are scarce~\cite{FePdN, FePdN1, FePdN2}. To the best of our knowledge, only Takahashi \textit{et al.} \cite{Thin_film} have reported on Pd-doped Fe$_4$N thin films, determining the elastic modulus of PdFe$_3$N to be 171\,GPa~in the ferrimagnetic state. Pd doping has also been recently explored in Co$_4$N, which shares structural similarities with Fe$_4$N. In Co$_4$N, Pd incorporation markedly increases the LP and enhances nitrogen retention without degrading magnetic properties \cite{Akshaya}. Therefore, Pd doping can be a promising strategy to further tune the electronic and magnetic properties of \fn~as evidenced in the present work.

In this work, we have studied the effect of Pd doping on the growth parameters such as nitrogen partial pressure (RN$_2$) and Ts. Also, we have varied atomic concentration of Pd while keeping the RN$_2$ and Ts fixed to investigate its effect on magnetic properties. Further, structural, magnetic and electronic properties of undoped and Pd doped samples have been studied.

\section{Experimental Details}

Fe$_4$N thin films were prepared using a direct current magnetron sputtering (dcMS) system (Orion-8, AJA Int. Inc.) on Si (100) and fused silica (SiO$_2$) substrate. A background pressure of $1\times10^{-7}$ Torr or lower was always achieved prior to deposition. The working pressure was typically maintained around $3\times10^{-3}$\,Torr due to combined flow of Ar and N$_2$ gases at 50\,sccm. A 20\,nm thick underlayer and 3\,nm thick capping layer of TiN was used to prevent substrate-film interdiffusion and surface oxidation. Fe$_4$N thin films were first optimized at Ts = 673\,K and RN$_2$ = 13\,\p. Pd doping was varied by applying a power of 2, 5 and 10\,W~resulting in Pd concentration of 5, 13, and 24~at.\,\p, respectively as determined from energy dispersive x-ray spectroscopy (EDS) measurements (not shown). The Pd doping was applied at a fixed RN$_2$ = 13\,\p~and at Ts~=~673\,K. Subsequently, RN$_2$ was varied from 10 to 16\,\p~while keeping the Ts at 673\,K and Pd at 5~at.\,\p. The thickness of the samples was in the range of 80-100\,nm. Furthermore, the impact of the Ts was explored by depositing Fe$_4$N films at 573 and 473\,K with RN$_2$ = 13\,\p, without and with 5 at.\p~Pd doping.

X-ray diffraction (XRD) and reflectivity (XRR) measurements were performed, respectively, using a Bruker D8 Advance or Discover diffractometer with Cu K$\alpha$ radiation ($\lambda$ = 1.54\,\AA).~Magnetic properties were examined using a Quantum Design SQUID-VSM (S-VSM) magnetometer.The local environment were studied using Fe K-edge x-ray absorption spectroscopy (XAS) measurement done at 8-ID ISS beamline (NSLS-II, USA)~\cite{EXAFS_ISS} and Fe L-edge at BL-01 (Indus 2, India)~\cite{BL01}.~Hard x-ray photoemission spectroscopy (HAXPES) was carried out at the P22 beamline of PETRA III (DESY, Germany) using 6\,keV incident photons to investigate the chemical bonding environment \cite{HAXPES_beamline}.~Secondary mass ion spectroscopy (SIMS) with a source of O$_2^+$ ions (5 \,keV, 400 \,nA) was utilized to sputter and obtain the chemical composition of the film using HIDEN Analytical SIMS Workstation.~This integrated approach enabled a detailed investigation of the influence of Pd doping and deposition temperature on the structural, magnetic, and compositional characteristics of Fe$_4$N thin films.

\section{Results and Discussion}
\subsection{Structural Characterization}
\begin{figure*}
    \centering
    \includegraphics[width=\textwidth, height=0.8\textheight, keepaspectratio]{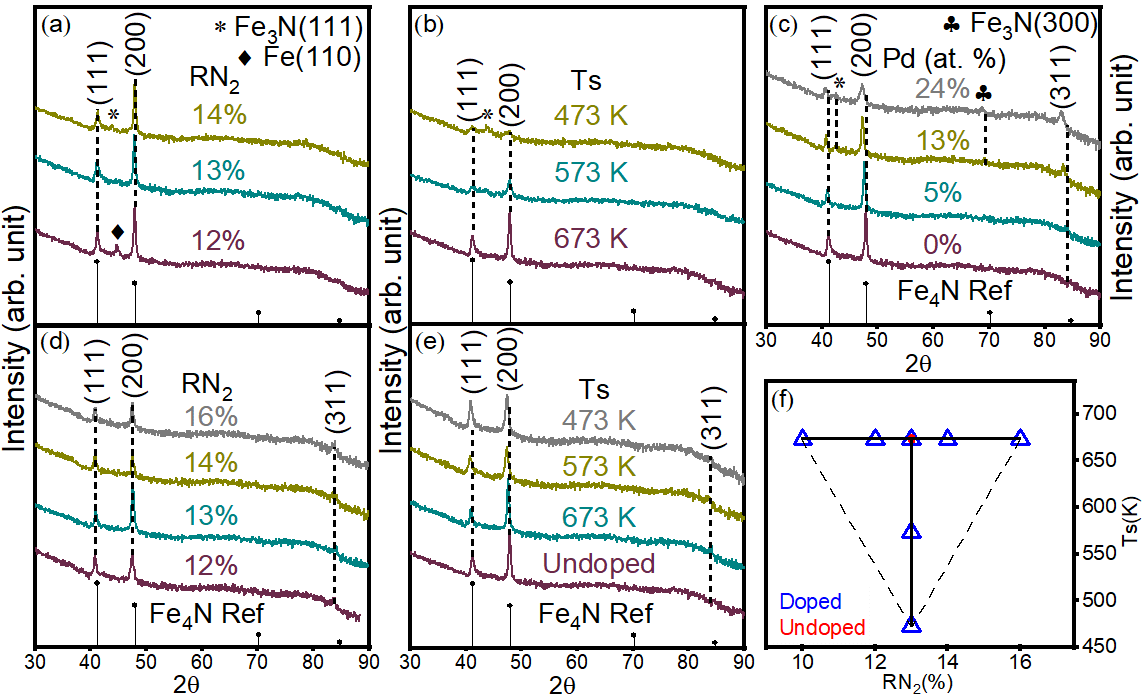}
    \caption{XRD pattern of Fe$_4$N thin films deposited on amorphous SiO$_2$ substrates at 673 K with RN$_2$ = 12, 13 and 14\,\p \,(a) and at RN$_2$ = 13\,\p  with Ts = 673, 573 and 473\,K (b) without doping. Further, XRD pattern of thin films at different Pd doping keeping RN$_2$ = 13\,\p\,and Ts = 673\,K (c). Extension of RN$_2$ range from 12-16\,\p\,at fixed Ts = 673\,K with doping (d) and comparison of undoped and Pd doped samples deposited at Ts = 673, 573, and 473\,K at a fixed RN$_2$ = 13\,\p (e). N concentration (RN$-2$) and growth temperature (Ts) diagram depicting the phase regime where the formation of Fe$_4$N phase takes place for undoped and Pd doped samples (f).}
    \label{XRD}
\end{figure*}

XRD pattern of samples corresponding to different growth parameters are shown in figure~\ref{XRD}.~To optimize a single-phase Fe$_4$N, the RN$_2$ was varied as 12, 13, and 14\,\p~at Ts~=~673\,K as shown in figure~\ref{XRD}(a). The XRD results indicate the presence of $\alpha$-Fe impurity peaks at RN$_2$ = 12\,\p~attributed to nitrogen deficiency and $\epsilon$-Fe$_3$N impurity phases at RN$_2$ = 14\,\p~due to nitrogen excess. A pure Fe$_4$N phase could be observed when RN$_2$ = 13\,\p, exhibiting (111) and (200) reflections at 2$\theta$ = 41.23 and 47.90\,\degree, respectively, consistent with JCPDS reference no. \#83-0875. The calculated values of the LP and average crystallite size (CS) were found to be 3.795~\AA~and 27.6\,nm, respectively, aligning well with reported theoretical [\cite{LatticeP}] and experimental [\cite{PANDEY201936, LP_exp,LP_expt,LP_expt1}] values for well established Fe$_4$N samples. Subsequently, to examine phase stability at lower  temperatures, the Ts was varied while maintaining RN$_2$ at 13\,\p~and resulting XRD patterns are shown in figure~\ref{XRD}(b). The emergence of $\epsilon$-Fe$_3$N impurity phases at the reduced Ts of 573 and 473\,K suggests a thermally unstable Fe$_4$N phase under these conditions, highlighting the importance of keeping the Ts fixed at 673\,K for the growth of Fe$_4$N phase.

It therefore becomes evident that Fe$_4$N has a very narrow Ts-Nc regime, characterized by RN$_2$ = 13\,\p~and Ts = 673\,K as shown by a red dot in figure~\ref{XRD}(f).~Therefore, Pd doping in Fe$_4$N was applied by varying its concentration at 5, 13 and 24\, at.\,\p~while keeping the RN$_2$ and Ts constant at 13\,\p~and 673\,K, respectively. The XRD pattern of these samples are compared in figure~\ref{XRD}(c). Here, it can be observed that a single phase Fe$_4$N is forming at 5\,at.\,\p~Pd but at higher Pd concentrations, peaks corresponding to $\epsilon$-Fe$_3$N phase can also be observed. Further, the broadening and shift to of peaks to lower angles with a rise in Pd doping can also be observed.~This broadening and shift implies that when Pd doping is increased beyond 5 at.\p, it create a disorder.~Therefore, 5\,at.\,\p~Pd doping seems to be best suited. From Table~\ref{LP&CS table}, increase in LP and decrease in CS can be clearly observed indicating the lattice expansion and increase in grain boundaries with increasing Pd doping.~Also, very small change in the LP at higher doping indicate the solubility limit of Pd into Fe$_4$N lattice. 

\begin{table}
\centering
\caption{Lattice parameter (LP) and crsytallite size (CS) of Fe$_4$N thin films obtained from analysis of XRD data with different Pd doping.}
\begin{tabular}{ccc} 
\hline
Pd (at.\,\p) & LP (\AA~$\pm$0.005) & CS (nm$\pm$ 1) \\
\hline
0   & 3.795 & 27  \\
5   & 3.815 & 23  \\
13  & 3.840 & 21  \\
24  & 3.844 & 11   \\
\hline
\end{tabular}
\label{LP&CS table}
\end{table}

Further, keeping the Pd concentration fixed at 5\,at.\,\p, the RN$_2$ and the Ts has been varied as shown in figure~\ref{XRD}(d) and~\ref{XRD}(e), respectively.~It can be seen here that the RN$_2$ range has now been extended to as low as 12\,\p~to as high as 16\,\p~in comparison to a fixed value of 13\,\p~for achieving the \fn~phase. Also, the Ts range is lowered down to 473\, K from 673\,K~indicating phase stability of \fn~at lower temperatures. A schematic representation of this extended Ts-Nc regime is shown in figure~\ref{XRD}(f). Here, the solid line denotes the region obtained experimentally, whereas the dashed line corresponds to the expected region for the growth of an optimum \fn~phase.

\subsection{Magnetic Properties:} 
\begin{figure}
    \centering
    \includegraphics[width=0.8\linewidth]{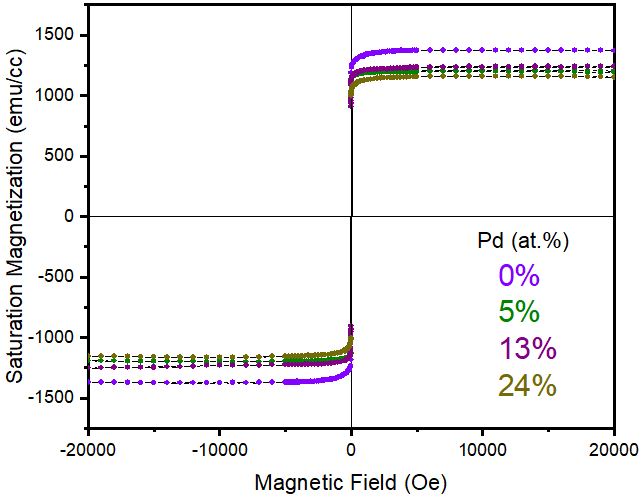}
    \caption{M-H loops showing the Ms values for undoped and 5, 13, 24 at.\,\p~Pd doped \fn~samples.}
    \label{VSM}
\end{figure}

Bulk magnetization measurement using S-VSM were performed to determine the Ms of undoped and Pd doped \fn~samples and the obtained M-H loops are shown in figure \ref{VSM}. The obtained values of the Ms are: 1375, 1202, 1244 and 1158 emu/cc, respectively for 0, 5, 13, and 24 at.\,\p~Pd doped \fn~samples. It can be noticed here that very less changes in Ms values is observed in 0, 5, and 13 at.\,\p~doped samples but it reduces significantly in the 24 at\,\p~Pd doped sample. At low Pd concentrations, Pd substitution introduces competing effects on the magnetic behavior. While Pd is non-magnetic and therefore reduces the net moment through magnetic dilution, its larger atomic size leads to lattice expansion. Further, resulting into reduction in Fe–Fe hybridization, thereby increasing the local Fe magnetic moments. This magneto volume-driven enhancement partially offsets the loss of magnetization caused by Pd substitution. However, at higher Pd concentrations, this compensation mechanism becomes ineffective.

\subsection{Electronic properties.} 
\begin{figure}
    \centering
    \includegraphics[width=0.8\linewidth]{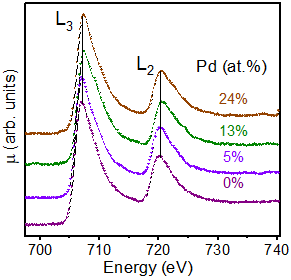}
    \caption{XANES spectra taken at Fe L$_3$ and L$_2$ edge for undoped and 5, 13, 24 at.\,\p~Pd doped \fn~samples.}
    \label{XAS}
\end{figure}

XAS measurement at Fe L-edge were performed for undoped and Pd doped samples in order to investigate the effect of doping on the oxidation state and local structure of the samples. Fe L-edge shows two main peaks L$_3$ (2$p$$_{\text{3/2}}$) and L$_2$ (2$p$$_{\text{1/2}}$) at photon energies of $\sim$707\,and $\sim$720\,eV due to well-known intrinsic spin-orbit coupling as shown in figure \ref{XAS}.~Pre-edge and post-edge background correction were performed to normalize and compare the data using Athena software package \cite{Athena}.~No change in undoped and 5\,at.\,\p~Pd doped Fe-L edge was observed, indicating that the local environment does not differ significantly with doping. Gradual shift in Fe L-edge peak structures is observed for 13 and 24\,at.\,\p~Pd doped samples towards higher energy which indicates increase in the oxidation state of Fe.~This shift in Fe L-edges can be understood from XRD (shown in figure~\ref{XRD}(c)) of undoped and  Pd doped samples as Fe$_4$N peaks intensity and shape started changing  at higher doping (24\,at.\,\p) and also peaks corresponding to Fe$_3$N appear. Fe$_3$N with higher oxidation state than Fe$_4$N can be the reason for gradual shift and change in oxidation state.

EXAFS measurements were performed on undoped and Pd doped Fe$_4$N samples to study the local structure and obtain information about the radial distance and the coordination number. Fe$_4$N exhibit anti perovskite type structure in which Fe atoms occupy the face centered (Fe II) and corner position (Fe I) and N atoms occupies the body centered position. The interatomic distance (R) between the Fe II - N and Fe II - Fe I atoms is approximately 1.89 and 2.68\,\AA, respectively. In case of Pd doping, distance between the Fe II and Pd atom is $\approx$ 2.72\,\AA. 

\begin{figure}
    \centering
    \includegraphics[width=0.8\linewidth]{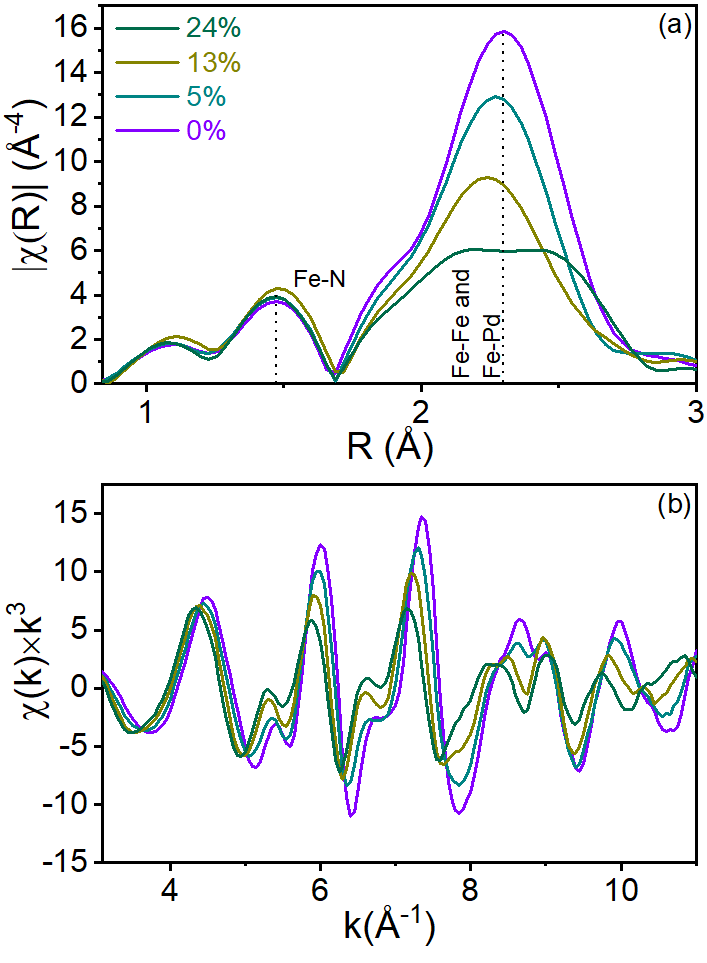}
    \caption{ The moduli of Fourier transform (FT) of Fe K-edge EXAFS (a) and $\chi$(k)$\times$k$_3$ spectra (b) of undoped and 5, 13, 24\,\p~Pd (at. Conc.) doped Fe$_4$N samples.}
    \label{raw exafs}
\end{figure}

Fourier transform (FT) of EXAFS and $\chi$(k){$\times$}k$^3$ spectra of undoped and 5, 13, 24\,at.\,\p~Pd doped Fe$_4$N samples shown in figure~\ref{raw exafs}. It can be observed from the figure~\ref{raw exafs}(a), FT modulus amplitude corresponding to Fe II - Fe I bonding is decreasing with increasing Pd concentration significantly while amplitude of Fe II - N is not showing much difference.~This decrease in the amplitude can be understood by decrease in the coordination number and increase in the disorder which are correlated with each other.~To further understand this, the $\chi$(k) spectra of undoped and Pd doped samples has been compared in figure~\ref{raw exafs}(b). It can be observed that $\chi$(k) amplitude show continuous decrement with increase in the Pd concentration from 0 to 24 at.\,\p~in higher and lower k region. This uniform decrement in the lower to higher k-range in $\chi$(k) is mainly related to decrease in the coordination number as reported by  Kumar \textit{et. al.} \cite{Yogesh}. 

\begin{figure*}
    \centering
    \includegraphics[width=0.75\linewidth]{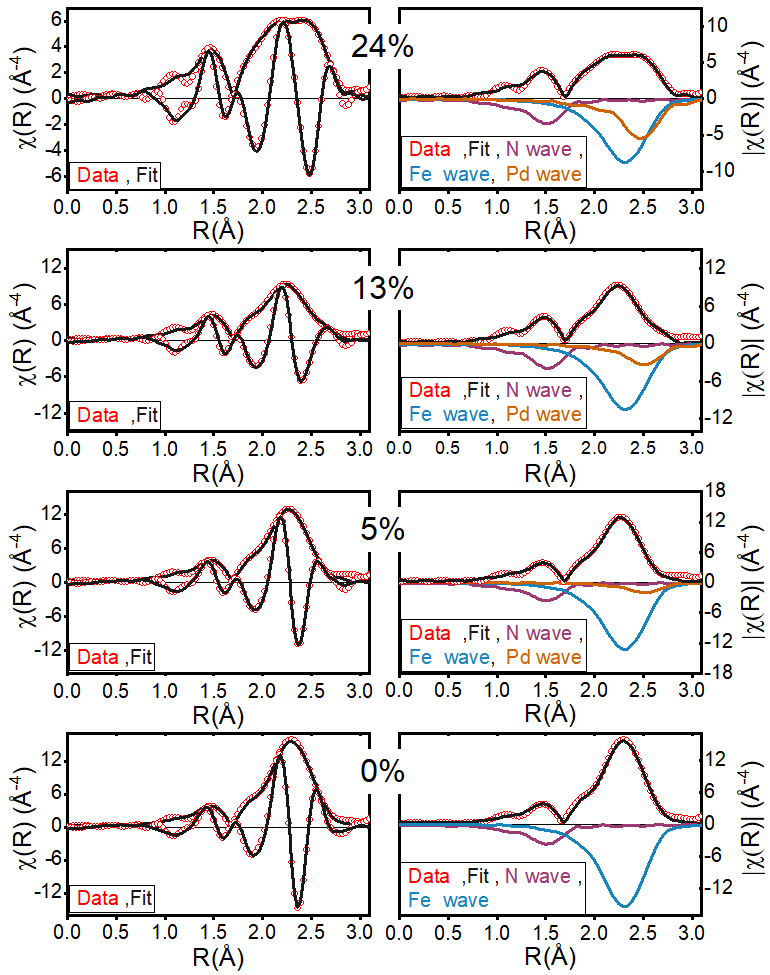}
     \caption{ Fourier transform moduli ($|\chi(R)|$) and real component (Re{$|\chi(R)|$}) of Fe K-edge EXAFS of undoped and Pd-doped Fe$_4$N samples, with raw data (red) and fit (black). Individual components used to fit spectra are shown inverted in the figure. The spectra are fitted with three paths: Fe--N, Fe--Fe , and Fe--Pd .}
    \label{EXAFS}
\end{figure*}

Quantitative analysis of the EXAFS data was performed by data fitting using the Artemis software package \cite{Athena} shown in figure~\ref{EXAFS}.~To ensure reliable fitting, all the results were constrained within acceptable parameter limits: R-factor < 2\,\p, $\Delta$ R < 0.5, $\Delta$ E$_0$ < 10, Debye–Waller factor ($\sigma$$^2$) < 0.03, and amplitude reduction factor (S$_0$$^2$) in the range of 0.7 to 1.05 \cite{EXAFS_constraints}. These criteria collectively ensured high quality EXAFS fits. The fitting was carried out in the R space within a range of 1 to 3 \AA.~The S$_0$$^2$ was fixed at 0.71 for all samples to maintain consistency across the dataset. For both undoped and doped samples, fitting models were constructed using scattering paths obtained from the corresponding $\textit{feff}$ files. Specifically, two single scattering paths Fe {II} – N and Fe {II} – Fe {I} were used for undoped samples, while an additional Fe {II} – Pd path was incorporated for the doped samples. Coordination number for Fe II - N path was kept fixed. Obtained parameters from the EXAFS fitting are listed in table~\ref{EXAFS table}. 

\begin{table*}[htbp]
\small
\centering
\caption{EXAFS metrical parameters obtained from the fittings of undoped and Pd-doped Fe$_4$N samples. $R$: atomic pair distance (\AA), N: coordination number, $\sigma^2$: Debye–Waller factor (\AA$^2$). Paths: Fe ${\text{II}}$--N (I), Fe ${\text{II}}$--Fe ${\text{I}}$ (II), Fe ${\text{II}}$--Pd (III).}
\vspace{0.5em}
\begin{tabular}{llllll}
\hline
{Path} & {Parameter} & {0\,\p} & \textbf{5\,\p} & {13\,\p} & {24\,\p} \\
\hline
\multirow{3}{*}{Fe ${\text{II}}$--N } 
& R              & 1.8978 & 1.9058 & 1.9133 & 1.9194 \\
& N              & 2      & 2      & 2      & 2 \\
& $\sigma^2$     & 0.0034 & 0.0036 & 0.0029 & 0.0037 \\
\hline
\multirow{3}{*}{Fe ${\text{II}}$--Fe ${\text{I}}$ } 
& R              & 2.6805 & 2.6885 & 2.7071 & 2.7116 \\
& N              & 12     & 11     & 10     & 8.4 \\
& $\sigma^2$     & 0.0101 & 0.0106 & 0.0115 & 0.0114 \\
\hline
\multirow{3}{*}{Fe ${\text{II}}$--Pd } 
& R              & --     & 2.7209 & 2.7094 & 2.6821 \\
& N              & --     & 1      & 2      & 3.6 \\
& $\sigma^2$     & --     & 0.0073 & 0.0083 & 0.0089 \\
\hline
\end{tabular}
\label{EXAFS table}
\end{table*}

It can clearly observed from the fitting of EXAFS data that the coordination number for Fe is decreasing with increasing Pd doping for Fe {II} - Fe {I} scattering path. While, for Fe {II} - Pd path coordination number is increasing with Pd doping which indicate the  replacement around Fe {II} atoms by Pd atoms. Further, increase in interatomic distance between for Fe {II} - N and Fe {II} - Fe I paths was observed with increasing Pd doping due to Pd incorporation, causing the lattice expansion as atomic radius of Pd is larger than Fe and N. While, interatomic distance for Fe {II} - Pd path is decreasing with increasing Pd doping. Also, increase in $\sigma$$^2$ was observed for Fe {II} - Pd path at higher doping which indicate the increase in the disorder due to the Pd incorporation. Increase in disorder can be related to replacement or clustering of Pd at Fe sites which may be causing structural rearrangement and lattice distortion around higher Pd doping. These local structural changes and lattice distortion may account for the appearance of Fe$_3$N phase in XRD and significant reduction in the Ms at 24 at.\,\p~Pd doping.

\begin{figure*}
    \centering
   \includegraphics[width=0.8\linewidth]{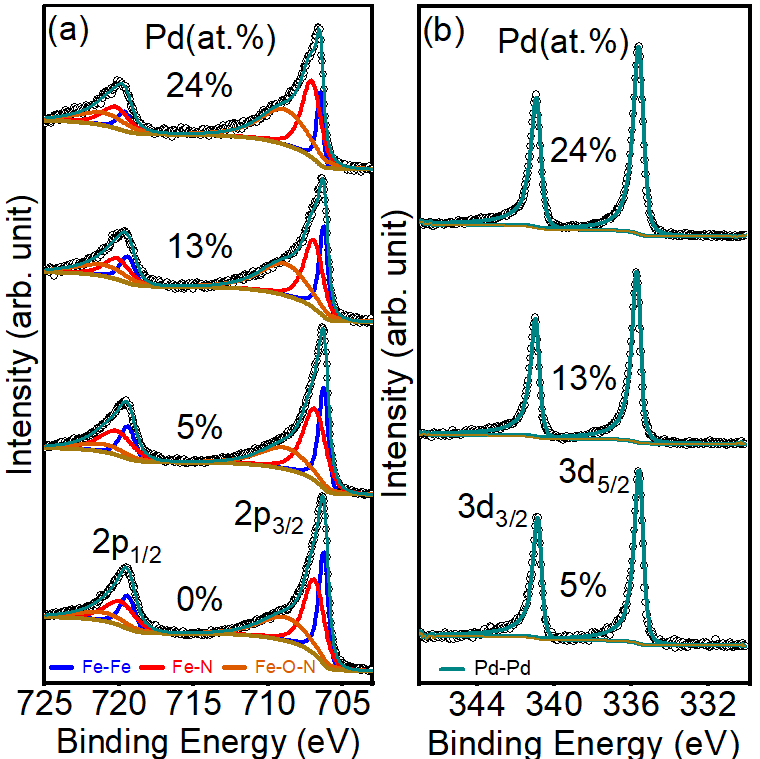}
    \caption{ HAXPES fitted spectra of undoped and 5, 13, 24\,\p~Pd doped Fe$_4$N for Fe-2p (a) and Pd-3d (b). }
    \label{XPS}
\end{figure*}

To investigate the chemical structure modifications induced by Pd substitution in Fe$_4$N, HAXPES measurement were performed on undoped and Pd doped Fe$_4$N (Pd = 5, 13, 24\,at.\,\p) samples.~The Fe 2$p$ core-level spectra (Figure~\ref{XPS}(a)) display well-resolved spin-orbit doublets corresponding to the Fe 2$p$$_{\text{3/2}}$ and 2$p$$_{\text{1/2}}$ levels.~Deconvolution of the Fe peaks reveal multiple chemically distinct components attributed to Fe–Fe metallic bonding (706.2 and 719.40\,eV), Fe–N covalent interactions (706.68 and 720.03\,eV), and Fe–O–N (708.36 and 720.69 eV) species due to oxidation of samples which are  matching very well with literature \cite{XPS1,XPS}. At higher Pd content, the Fe–N peak becomes more prominent, accompanied by a suppression of the Fe–Fe feature, indicating a transition toward a more nitrogen rich local environment. Importantly, we observe a slight shift in the Fe 2$p$ peaks toward higher binding energy at 24 at.\,\p~Pd doping i.e. higher oxidation state complimenting the Fe-L edge XAS and XRD results.~The Pd 3$d$ core-level spectra shown in figure~\ref{XPS}(b). The characteristic Pd 3$d$$_{\text{5/2}}$ and 3$d$$_{\text{3/2}}$ peaks are observed near 335.0 and 340.3\,eV, respectively. While the line shape remains dominated by metallic Pd$^0$ states. Notably, no spectral signatures of PdO or PdN were detected, affirming the chemical stability of Pd$^0$ within the matrix.

\section{Conclusion}
Single-phase Fe$_4$N thin films were synthesized using dc magnetron reactive sputtering, both with and without Pd doping. The introduction of 5 at.\,\p~Pd significantly expanded the phase formation window, enabling the stabilization of single-phase Fe$_4$N over a broader RN$_2$ range and at reduced Ts. This indicates that Pd doping effectively extends the Ts-Nc regime, likely by enhancing nitrogen retention and stabilizing the crystal structure. Importantly, magnetic measurements revealed only a marginal decrease in saturation magnetization for the Pd-doped sample compared to the undoped counterpart, suggesting that the intrinsic magnetic properties of Fe$_4$N are largely preserved. These results highlight the potential of Pd as a useful dopant for tailoring the synthesis conditions and phase stability of Fe$_4$N without significantly compromising its magnetic performance.
\section*{Acknowledgments}
This research was support by junior research fellowship awarded by UGC, Government of India to RKM. Portions of this research were carried out at the light source PETRA - III of DESY, a member of Helholtz Association (HGF). Financial support by the Department of Science \& Technology (Government of India) provided within frame work of India{$@$}DESY collaboration is gratefully acknowledged. Experimental support received from R.J. Choudhary, R. Mishra, R.\,Sah and L.\,Behera is gratefully acknowledged.  

\bibliography{Reference}
\end{document}